\documentclass[aps,pra,reprint,superscriptaddress]{revtex4-1}
\usepackage{amsmath}
\usepackage{amssymb}
\usepackage{graphicx}
\usepackage[caption=false]{subfig}
\usepackage{enumitem}
\usepackage{color}
\usepackage[percent]{overpic}
\usepackage[normalem]{ulem}

\newcommand{\ket}[1]{{\big| {#1} \big>}}
\newcommand{\bra}[1]{{\big< {#1} \big|}}

\DeclareMathOperator{\tr}{tr}
\newcommand{\ii}{\mathrm{i}}

\begin{document}

\title{Perfect Zeno-like effect through imperfect measurements at a finite frequency}

\author{David Layden}
\affiliation{Department of Applied Mathematics, University of Waterloo, Waterloo, Ontario, N2L 3G1, Canada}
\affiliation{Institute for Quantum Computing, University of Waterloo, Waterloo, Ontario, N2L 3G1, Canada}
\author{Eduardo Mart\'{i}n-Mart\'{i}nez}
\affiliation{Department of Applied Mathematics, University of Waterloo, Waterloo, Ontario, N2L 3G1, Canada}
\affiliation{Institute for Quantum Computing, University of Waterloo, Waterloo, Ontario, N2L 3G1, Canada}
\affiliation{Perimeter Institute for Theoretical Physics, 31 Caroline St N, Waterloo, Ontario, N2L 2Y5, Canada}
\author{Achim Kempf}
\affiliation{Department of Applied Mathematics, University of Waterloo, Waterloo, Ontario, N2L 3G1, Canada}
\affiliation{Institute for Quantum Computing, University of Waterloo, Waterloo, Ontario, N2L 3G1, Canada}
\affiliation{Perimeter Institute for Theoretical Physics, 31 Caroline St N, Waterloo, Ontario, N2L 2Y5, Canada}

%\date{\today}

\begin{abstract}
The quantum Zeno effect is usually thought to require infinitely frequent and perfect projective measurements to freeze the dynamics of quantum states. We show that perfect freezing of quantum states can also be achieved by more realistic non-projective measurements performed at  a finite frequency. 
\end{abstract}


\maketitle

%%%%%%%%%%%%%%%%%%%%%%%%%%%%%%%%%%%%%%%%%%%%%
\section{Introduction}
The quantum Zeno effect (QZE), i.e., the phenomenon that the dynamics of a quantum system can be inhibited by frequent measurements, has recently attracted increased interest as a tool in the field of quantum information processing (QIP).  There, the QZE promises a wide range of applications, potentially playing a role in error-correcting codes \cite{Erez:2004, Silva:2012, Dominy:2013}, decoherence-free subspaces \cite{Beige:2000}, entanglement production and state preparation \cite{Nakazato:2003, Nakazato:2004, Wang:2008}, as well as gate implementation \cite{Franson:2004}. Moreover, due to experimental advances in the last decade, there has also been a growing interest in Zeno-like effects in the context of light-matter interactions \cite{Beige:2000, Maniscalco:2008, Bernu:2008, Raimond:2012}.

In the conventional QZE, repeated instantaneous perfect measurements are performed on the system. In the limit of infinite measurement frequency, and only in that limit, does the evolution of the measurement's eigenstate freeze. To date, experimental efforts have focused, therefore, on trying to approximate as best as possible the challenging limit of  infinitely-frequent instantaneous perfect measurements \cite{Itano:1990, Fischer:2001,Xiao:2006, Bernu:2008, Zheng:2013}.

Here we show that realistic, i.e., imperfect measurements, performed at a finite frequency, can also completely freeze the state of a system.  In this sense, we can achieve a perfect Zeno-like effect with finite-frequency imperfect measurements. Moreover, which state of the system will be frozen can be chosen by controlling parameters such as the strength and frequency of the repeated imperfect measurements. In contrast, in the usual Zeno effect (that follows from repeated instantaneous perfect measurements at infinite frequency) there is much less choice, as the state to be preserved can only be an eigenstate of the repeatedly measured observable. 

As a technical tool, in order to be able to determine to what extent realistic finite-duration imperfect measurements can freeze the dynamics, we will enlarge the Heisenberg cut so as to include also the measurement devices in the quantum description \cite{Petrosky:1990, Peres:1990, Pascazio:1994, Ruseckas:2001, Facchi:2002, Ai:2013}. We will call the system under consideration the target, and the measurement devices will be called detectors. The measurements will be described by a suitable target-detector interaction Hamiltonian, and the repeated measurement processes will each be unitary. The derivation of our results will not invoke any projective measurements or wavefunction collapse. 

%{\blu After the detector is decoupled one may then read out the detector and/or let the detector interact with an environment. With our location of the Heisenberg cut, it is at this stage, that classical outcomes of the detector readout will be generated with classical probabilities.}

%%%%%%%%%%%%%%%%%%%%%%%%%%%%%%%%%%%%%%%%%%
\section{Setting}

The target and detector are initially uncorrelated, i.e., they are in a product state.  We let the target system evolve alone under its free Hamiltonian $H_0^{(\textsc{t})}$ for a time interval $\Delta t_\textsc{f}$.  We then measure the target by coupling it to a first detector, which is prepared in the state $\rho_{\, 0} ^{(\textsc{d})}$.  We let the bipartite system evolve under the Hamiltonian $H_\textsc{m} = H_\textsc{i} + H_0^{(\textsc{t})} + H_0^{(\textsc{d})}$ for a time $\Delta t_\textsc{m}$, where $H_\textsc{i}$ is the interaction Hamiltonian describing the coupling and $H_0^{(\textsc{d})}$ is the free Hamiltonian of the detector.  After the measurement has taken place, we decouple the systems, set aside the detector, and repeat the process with a fresh detector.  One may either use a series of identically-prepared detectors, or equivalently, one may use a single detector which is reset after each interaction.

The net effect of each free evolution plus measurement cycle on the target system can be described as a quantum channel, $\Phi: \mathcal{L}(\mathcal{H}_\textsc{t}) \rightarrow \mathcal{L}(\mathcal{H}_\textsc{t})$, where $\mathcal{L}(\mathcal{H}_\textsc{t})$ denotes the set of linear maps on the target's Hilbert space.  To ascertain the effect of many such cycles on target states, we analyze the channel's spectrum, $\{\lambda_j \} \subset \mathbb{C}$, and its eigenvectors $\{V_j \} \subset \mathcal{L}(\mathcal{H}_\textsc{t})$.  While the $V_j$'s need not correspond to physical states when considered individually, any initial target state can be decomposed into the eigenbasis which they form, as $\rho_{\, 0} ^{(\textsc{t})} = \sum_j c_j V_j$. Note that this method readily generalizes to the case where $\Phi$ is defective since the latter still admits a Jordan Canonical Form.  After $n$ cycles, the target state will be
\begin{equation}\label{eq:channel}
\rho_{\, n} ^{(\textsc{t})} = \Phi^{n} \big(\rho_{\, 0} ^{(\textsc{t})} \big) = \sum_j \lambda_j^n c_j V_j,
\end{equation}%
%`
and thus components of $\rho_{\, 0} ^{(\textsc{t})}$ in eigenspaces with $\lambda = 1$ will be preserved, while those in $|\lambda| < 1$ eigenspaces are exponentially suppressed.

Fixed points of the process, for which $\Phi^{n} \big( \rho_{\, 0} ^{(\textsc{t})} \big) = \rho_{\, 0} ^{(\textsc{t})}$ for all $n$, are those that lie entirely in a $\lambda = 1$ eigenspace of $\Phi$.  We can straightforwardly apply Brouwer's fixed point theorem \cite{Brouwer} here to conclude that such a state always exists.  If the decomposition of a given state $\rho_{\, 0} ^{(\textsc{t})}$  into eigenspaces of $\Phi$ involves only eigenspaces whose eigenvalues are very close to unity, then the exponential suppression of $|\lambda| < 1$ eigenvalues can be exceedingly slow. In that case, $\Phi^{n} \big( \rho_{\, 0} ^{(\textsc{t})} \big) \approx \rho_{\, 0} ^{(\textsc{t})}$, for values of $n$ which are not too large.  If an eigenvalue $\lambda_j$ is not equal to but close to unity, let us refer to its corresponding eigenvector $V_j$ as an `almost fixed point' of $\Phi$.   We will show that two types of fixed points and the `almost fixed points' are at the heart of the Zeno phenomena. 

%%%%%%%%%%%%%%%%%%%%%%%%%%%%%%%%%%%%%%%%%%%%%
\section{A qubit model}
To demonstrate the emergence of the QZE from imperfect measurements in a manner that is simple and  experimentally relevant, we consider the case in which the target and detectors are qubits.

Without loss of generality, we take the free Hamiltonian of the target qubit to be $H_0^{(\textsc{t})} = \omega \sigma_z^{(\textsc{t})}$.  We then consider a Jaynes-Cummings-like coupling \cite{Scully:1997}
\begin{equation} \label{eq:H_int}
H_\textsc{i} = g \sigma_x^{(\textsc{d})} \sigma_x ^{(\textsc{t})}
	= \frac{g}{4} \Big( \sigma_+^{(\textsc{d})} +  \sigma_-^{(\textsc{d})}\Big) \Big( \sigma_+^{(\textsc{t})} +  \sigma_-^{(\textsc{t})}  \Big).
\end{equation}
For simplicity, we take the free Hamiltonian of the detectors to have the same form as that of the target, namely $H_0^{(\textsc{d})} = \omega \sigma_z^{(\textsc{d})}$. 
In order to be explicit, for now we choose the detectors to arrive in their ground state, i.e., $\rho_{\, 0} ^{(\textsc{d})} = |0\rangle_\textsc{d} \langle{0}|_\textsc{d}$. (Further below, we will discuss the implications of choosing an arbitrary other initial state for the detectors).  While our results will hold for more general choices, this choice of coupling and initial detector state commonly appears in real-world systems, e.g., in circuit QED \cite{degroot:2010, Majer:2007} and dipole-dipole coupling \cite{Ficek:2004} Ch.\ 16, and thus provides a physically relevant example.

%%%%%%%%%%%%%%%%%%%%%%%%%%%%%%%%%%%%%%%%%%
\subsection{Recovering the conventional quantum Zeno effect as a special case}
\label{sec:idealized}
To demonstrate the emergence of the usual Zeno phenomenology from our model, we begin by considering the limit of infinitely frequent measurements. In the usual description of the QZE, the target system is observed periodically, and the operators describing the measurement process do not depend on the measurement frequency (i.e., the duration of the measurements is assumed to be irrelevant).  To mimic this setup, we consider the limit $\Delta t_\textsc{f},\, \Delta t_\textsc{m} \rightarrow 0$ (infinite measurement frequency), but hold the strength of the interaction---described by $g \Delta t_\textsc{m}$---fixed by letting $g$ grow unbounded.

In this idealized setting, the eigenbasis $\{V_j\}$ of the channel $\Phi$ describing a single free evolution/measurement cycle, consists of the identity and the Pauli matrices.  As per Eq.~(\ref{eq:channel}), if the target qubit starts in the state  $\rho_{\, 0} ^{(\textsc{t})} = \frac{1}{2}(I^\textsc{(t)} + \boldsymbol{r} \cdot \boldsymbol{\sigma}^\textsc{(t)})$, then after $n$ cycles its state is
\begin{equation}
\rho_{\, n} ^{(\textsc{t})} = \frac{1}{2} \Big(
\lambda_0^n I^\textsc{(t)} + \sum_{j=1}^3 \lambda_j^n r_j \sigma_j^\textsc{(t)} \Big),
\end{equation}%
where 
\begin{align}
\lambda_0 &= 1\\
\label{eq:lambda_1}\lambda_1 &= 1- 2\omega^2\cot^2(g \Delta t_\textsc{m}) \Delta t_\textsc{f}^2 + \mathcal{O}(\Delta t_\textsc{f}^4) \\
\lambda_2 &= \lambda_3 = \cos(2 g \Delta t_\textsc{m}).
\end{align}%
We recall that to reproduce the usual QZE setting, $g \Delta t_\textsc{m}$ is finite and otherwise unconstrained, while $\Delta t_\textsc{f}$ is vanishingly small.  Thus, the expansion of $\lambda_1$ in Eq.~(\ref{eq:lambda_1}) is well-defined, as the values of $g \Delta t_\textsc{m}$ which cause the cotangent to diverge correspond to a trivial interaction.

We note the main features of this result: (i) The maximally mixed state is a fixed point of this process, as one might expect on the basis of decoherence-like effects produced by frequent interactions. (ii) The $\sigma_y^\textsc{(t)}$ and $\sigma_z^\textsc{(t)}$ components of the initial target state are exponentially suppressed, as $|\lambda_2| = |\lambda_3| < 1$ almost everywhere in the space of parameters $(g, \omega, \Delta t_\textsc{f}, \Delta t_\textsc{m})$. (iii) The eigenvalue $\lambda_1$ approaches unity quadratically in the high measurement frequency limit, so target states of the form $\rho_{\, 0} ^{(\textsc{t})} = \frac{1}{2}(I^\textsc{(t)} + r_1 \sigma_x^\textsc{(t)})$ are exactly preserved by strong and infinitely frequent measurements, even though they are affected by the free evolution.  This is an archetypal instance of the quantum Zeno effect.

(iv) In the current setting, the preservation of target states diagonal in the $\mathcal{X} = \{ \ket{0} \pm \ket{1} \}$ basis is the phenomenon typically associated with a QZE induced through frequent $\sigma_x ^{(\textsc{t})}$ projective measurements.  One might expect that, in this limit, the interaction constitutes a measurement of the $\sigma_x^{(\textsc{t})}$ component of the target's state.  Indeed, we observe that if $g\Delta t_\textsc{m} = \pi(2k+1) / 4$ for integer $k$, a general target state is mapped as
\begin{equation} \label{eq:state_update}
\begin{pmatrix}
a & b\\ 
b^* & c
\end{pmatrix}_{[\mathcal{X}]}
\mapsto
\begin{pmatrix}
a & 0\\0 & c
\end{pmatrix}_{[\mathcal{X}]}
\end{equation}
by the measurement, in the $\mathcal{X}$ basis.  In other words, any coherent superposition is collapsed to a probabilistic ensemble of $\sigma_x^{(\textsc{t})}$ eigenstates; exactly the effect of a $\sigma_x^{(\textsc{t})}$ projective measurement (PVM). Moving away from the above values of $g\Delta t_\textsc{m}$ leads to weaker $x$ measurements.

(v) We see that, remarkably, the wavefunction collapse in Eq.~(\ref{eq:state_update}) is not necessary to produce the ordinary QZE: As one increases the measurement frequency ($\Delta t_\textsc{f}\rightarrow0$), Eq.~\eqref{eq:lambda_1} shows the rate at which the ordinary Zeno effect is approached. We notice then that values of $g\Delta t_\textsc{m}$ can be chosen such that the term $\mathcal{O}(\Delta t_\textsc{f}^2)$ in \eqref{eq:lambda_1} vanishes, which implies that $\lambda_1$ will approach unity two powers in $\Delta t_\textsc{f}$ faster. Interestingly, the values of $g\Delta t_\textsc{m}$ that accomplish this do not correspond to the case of PVMs, but instead correspond to `weaker', effective POVM measurements.

%%%%%%%%%%%%%%%%%%%%%%%%%%%%%%%%%%%%%%%%%%
\subsection{Realistic Settings}
\label{sec:realistic}
In experiments, there are bounds on the frequency with which measurements can be repeated and on the coupling strengths between realistic detector and target systems.  Let us, therefore, go beyond the usual infinite frequency assumption into more experimentally-relevant regimes. As in the infinite-frequency case above we will analyze the eigenpairs $\{V_j', \lambda_j' \}$ of the channel $\Phi$ that represents a cycle---now at a finite measurement frequency and at finite coupling strength. We use primes to distinguish eigenvectors and eigenvalues from their infinite-frequency counterparts. 

We find that the eigenvalues and eigenvectors of $\Phi$ are now more involved than in the idealized case considered earlier. Concretely, the eigenvectors have the form
\begin{align}\label{eq:eigenvectors_prime}
V_0' &= \begin{pmatrix}
a & 0\\
0 & 1-a
\end{pmatrix}_{[\mathcal{Z}]} &
V_1' &= \begin{pmatrix}
0 & \theta\\
\varphi & 0
\end{pmatrix}_{[\mathcal{Z}]}\\
V_2' &= \begin{pmatrix}
0 & \chi \\
\eta & 0
\end{pmatrix}_{[\mathcal{Z}]} &
V_3' &= \begin{pmatrix}
b & 0 \\
0 & -c
\end{pmatrix}_{[\mathcal{Z}]} \nonumber
\end{align}
expressed in the $\mathcal{Z} = \{ \ket{0},\, \ket{1} \}$ basis, where $a$, $b$ and $c$ are non-negative and all other variables are complex.  The full form of these eigenvectors is discussed in the Appendix. Two of the associated eigenvalues are
\begin{align}
\lambda_0' &= 1\\
\label{eq:lambda_3_prime}\lambda_3' &= \cos^2(g \Delta t_\textsc{m}) -\frac{g}{\Omega} \sin^2 (\Omega \Delta t_\textsc{m}), 
\end{align}
where $\Omega = \sqrt{g^2 + 4\omega^2}$.  The expressions for $\lambda_1'$ and $\lambda_2'$ are lengthy. Their behavior is discussed below, and their full form is also discussed in the Appendix.  We note that, as in the previous subsection, while any $\rho_{\, 0} ^{(\textsc{t})}$ can be expressed in the eigenbasis $\{ V_j' \}$ of $\Phi$, only $V_0'$ represents a realizable state when considered alone.

With the exception of $\lambda_0'$, all of the variables in Eqs.~(\ref{eq:eigenvectors_prime}) -- (\ref{eq:lambda_3_prime}) depend on the parameters $(g, \omega, \Delta t_\textsc{f}, \Delta t_\textsc{m})$, or a subset thereof. We present the emergent features analogously to how we discussed the features of infinite measurement frequency case:

(a) The eigenvalue $\lambda_0' = 1$ is unique in that it is equal to unity independently of all parameters, making $V_0'$ a fixed point of $\Phi$.  The $\sigma_x^{(\textsc{d})} \! \otimes \sigma_x ^{(\textsc{t})}$ coupling in our scheme, together with the initial detector state, has the effect of rotating the state of the target qubit about the $x$-axis (on the Bloch sphere) during measurement.  However, from the symmetry of the scheme, target states which are diagonal in the $\mathcal{Z}$ basis are not preferentially rotated in either direction, and so they necessarily remain diagonal in this basis after the interaction.  Since these states form a compact, convex subset of a real vector space, Brouwer's theorem guarantees a fixed point which is diagonal in the $\mathcal{Z}$ basis; namely $V_0'$. Contrary to what intuition might suggest, $V_0'$ need not be the maximally mixed state for finite measurement frequencies. Rather, for the case where the detectors are prepared in the ground state, $V_0'$ can be anywhere on the $z$-axis of the Bloch sphere depending on the parameters.

(b) Generically, $|\lambda_3'|$ is smaller than unity, but as Eq.~(\ref{eq:lambda_3_prime}) shows, it can reach unity at isolated parameter values. If  $|\lambda_3'|<1$, it corresponds to a an exponentially damped transient. If $|\lambda_3'|=1$ then it merely contributes to the $z$-axis polarization of the fixed point.    

(c) $V_1'$ and $V_2'$ describe the off-diagonal elements of $\rho_{\, 0} ^{(\textsc{t})}$ in the $\mathcal{Z}$ basis, and in the limit of strong and high frequency measurements, they tend towards $\sigma_x^{(\textsc{t})}$ and $\sigma_y^{(\textsc{t})}$ respectively, as per the previous subsection. In this limit, $\lambda_1'$ tends towards unity, conserving the $\sigma_x^{(\textsc{t})}$ component of the initial target state.
From \eqref{eq:H_int}, this is indeed the limit of instantaneous $\sigma_x^{(\textsc{t})}$ measurements, i.e., we here recover the conventional Zeno effect. 
\begin{figure*}
\subfloat{
\raisebox{0em}{
\includegraphics[width=0.35\textwidth]{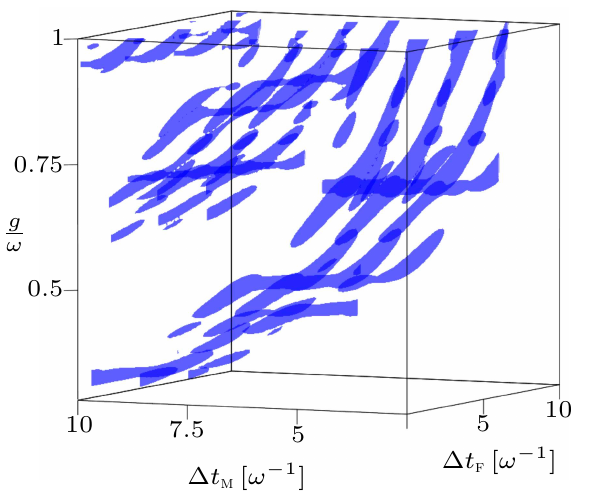}
\label{fig:1a}
}} \hspace{-2.1em}
\subfloat{
\raisebox{0em}{
\includegraphics[width = 0.33\textwidth]{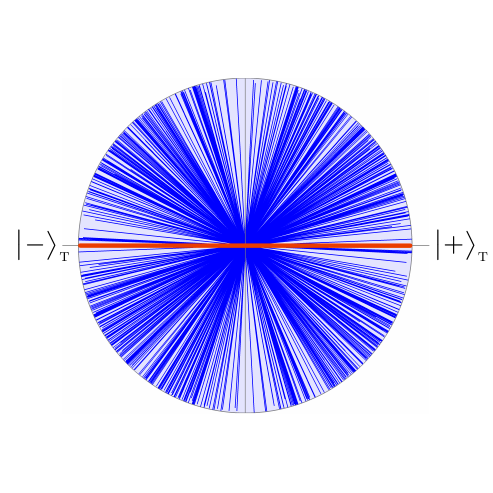}
\label{fig:1b}
}} \hspace{-1.5em}
\subfloat{
\raisebox{0em}{
\includegraphics[width=0.32\textwidth]{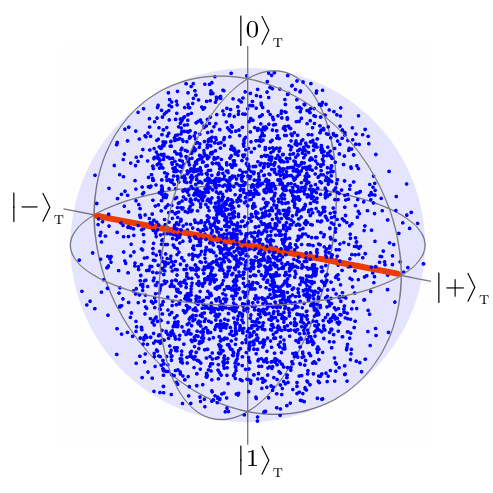}
\label{fig:1c}
}}
\caption{(Color online) \textbf{a)} Location of the Zeno-like fixed points in the 3-dimensional parameter space. Plotted are the neighborhoods in which the eigenvalue $\lambda_2'$ differs from 1 by at most $10^{-2}$. The thickness of these neighborhoods illustrates the level of fine-tuning required to realize Zeno-like fixed points with this accuracy. \textbf{b)} Top view of the Bloch sphere. Locations of Zeno-like fixed point states (where $\lambda_1' = 1$ or $\lambda_2' = 1$) are in blue. They are highly concentrated on the $xy$ plane. States preserved by the conventional Zeno effect are shown as a horizontal bold red line. \textbf{c)} Bloch sphere locations of target states preserved by Brouwer fixed points ($\lambda_0 = 1$) when the initial state of the detectors is arbitrarily varied, while being held fixed for each set of repeated measurements. We see that by varying the choice for the initial state of the detectors, the fixed points cover the full sphere isotropically about the $z$-axis.  For comparison, the bold red line on the $x$-axis shows the states that can be preserved by the conventional QZE.}
\end{figure*}
We also recover the phenomenon that, at finite measurement frequency,  the QZE is transient \cite{Itano:1990, Fischer:2001,Xiao:2006, Bernu:2008}. Namely, for high---but finite---measurement frequencies, generically $|\lambda_1'| \lesssim 1$, and therefore any mixture of  $\sigma_x^{(\textsc{t})}$ eigenstates is at best an `almost fixed point'. Furthermore, in our generalized framework we are now able to tell where the escape from the QZE will take the system: 
For an initial detector state $\rho_{\, 0} ^{(\textsc{d})} = |0\rangle_\textsc{d} \langle{0}|_\textsc{d}$,  the off-diagonal elements (in the $\mathcal{Z}$ basis) of an initial target state decay exponentially. In fact, more precisely, any target state will decay to the fixed point $V_0'$ under repeated measurement.

Crucially now, under closer inspection, we find that even though $|\lambda_1'|$ and $|\lambda_2'|$ are generically smaller than one, there exist  parameter values where $\lambda_1'$ or $\lambda_2'$ become arbitrarily close to, or reach unity. This occurs for a large number of parameter combinations corresponding to \emph{finite frequency} measurements with \emph{finite coupling strengths}, see Fig.~\ref{fig:1a}.  In other words, finite frequency imperfect measurements can preserve certain quantum states which are not preserved under free evolution; a phenomenon which we refer to as a ``Zeno-like effect".  

As Fig.~\ref{fig:1b} shows, the Zeno-like fixed points cover the entire $xy$ plane. This is a further improvement over the standard Zeno effect, where only states on the $x$-axis can be frozen by an interaction that measure $\sigma_x^{(\textsc{t})}$. We notice that the Zeno-like fixed points lie on the plane perpendicular to the polarization of the states of the incoming detectors. For generic detector polarizations the distribution of the Zeno-like fixed points can be moved out of the $xy$ plane. 
Even further, while Zeno-like fixed points only arise for specific choices of the parameters, strong results can also be shown for the `Brouwer' type fixed point $V_0'$ whose existence is guaranteed by Brouwer's theorem for all values of the parameters. Namely,  $V_0'$ can be made to lie anywhere on the $z$-axis of the Bloch sphere, which again happens to be the polarization axis of the initial state of the detectors. 

Then, crucially, as Fig.~\ref{fig:1c} shows, by choosing a suitable initial state for all the detectors in the series of measurements, we can make $V_0'$ be at any position in the Bloch sphere. This is a dramatic improvement over the standard Zeno effect since it implies that one can completely and stably halt the evolution of any state of the target system by suitably choosing interaction parameters and the initial state of the detector. In fact, for generic parameter values, i.e., in the absence of a Zeno-like fixed point, any state of the target system will eventually be driven towards $V_0'$, a state which one can choose arbitrarily by suitably choosing the initial state of the detectors and the measurement parameters.

\section{The target-detector interaction as a measurement}

Recall that we are including the detector inside the Heisenberg cut. This means that each detection process is described as the unitary evolution of a closed system. Our results about the Zeno-like effect do not depend on whether the detectors, once discarded, are read out or interact with an environment. Notice that this also means that, unlike in the usual QZE, in our approach one naturally does not post-select for measurement outcomes.

Nevertheless, for illustration, let us now consider the case when the discarded detectors are being read out in a specified way. Recall that, in this case, classical detection outcomes will be generated, occurring with classical probabilities. The full process can be described as a POVM. 

This POVM is determined by the requirement that it produces the correct probabilities for the outcomes of the detector readout, along with the correct state update rules for the target. Namely, this means that the Kraus operators are determined by the requirement that the quantum channel for the target is reproduced, and that the POVM elements correspond to the chosen readout variable.

Concretely, the channel $\Phi$ admits an operator-sum representation of the form
\begin{equation}
\Phi \big( \rho_{\, 0} ^{(\textsc{t})} \big)
= \sum_j
K_j \rho_{\, 0} ^{(\textsc{t})} K_j^\dagger,
\label{eq:kraus}
\end{equation}
where $\{K_j\} \subset \mathcal{L}(\mathcal{H}_\textsc{t})$ are Kraus operators. Defining $E_j = K_j^\dagger K_j$, it follows immediately that $\sum_j E_j = I^{(\textsc{t})}$, $E_j^\dagger = E_j$, and $E_j \ge 0$ (see \cite{nielsen&chuang} Ch.\ 2 \& 8). Thus, $\{E_j\}$ forms a POVM, with the probability of outcome $j$ given by
\begin{equation}
p_j = \tr \left( E_j \rho_{\, 0} ^{(\textsc{t})} \right).
\label{eq:probabilities}
\end{equation}
The post-measurement state corresponding to this outcome is
\begin{equation}
\rho_{\, 1,j} ^{(\textsc{t})} = \frac{K_j \rho_{\, 0} ^{(\textsc{t})} K_j^\dagger}{p_j},
\end{equation}
and so the action of the channel in Eq.~\eqref{eq:kraus} can be understood as
\begin{equation}
\Phi \big( \rho_{\, 0} ^{(\textsc{t})} \big) = 
\sum_j
p_j \, \rho_{\, 1,j} ^{(\textsc{t})}.
\label{eq:POVM_update}
\end{equation}
Eq.~\eqref{eq:POVM_update} shows that, as expected, the measurement process collapses the target to a classical ensemble of the different possible measurement outcomes. 

%{\red We note that the probabilities $\{ p_j\}$ in Eqs.~\eqref{eq:probabilities}--\eqref{eq:POVM_update} do not figure elsewhere in the current work, as the effects discussed above are entirely due to measurement back-action on the target system. Thus, the fact that there may exist different forms of Eq.~\eqref{eq:POVM_update} due to the non-uniqueness of Kraus decompositions is inconsequential here, as our scheme does not depend on the outcomes of the measurements.}

%We see in Eq.~\eqref{eq:POVM_update} {\color{red} that, in general, the post-measurement detector state is not mapped to an eigenstate of the measured observable. Instead, the measurement process drives the target to a classical ensemble of the different possible measurement outcomes of the POVM. This should not be surprising, since it is an immediate consequence of working with an expanded Heisenberg cut: Since we are describing the interaction between detector and target quantum-mechanically, the detector `collapses' the target state by introducing decoherence in the target system and therefore driving it to a classical ensemble of all possible outcomes.}

%that a general detector state is not mapped to the post-measurement state of a single outcome, but rather, to a probabilistic ensemble of states corresponding to all outcomes. This feature is an immediate consequence of working with a raised Heisenberg cut.

\subsection{An idealized example}

In the case of infinitely frequent projective measurements discussed in Section \ref{sec:idealized},  point (iv), we can explicitly compute a decomposition of $\Phi$ given by Eq.~\eqref{eq:kraus}:
\begin{equation}
K_+ = \ket{+}\bra{+} 
\qquad
K_- = \ket{-}\bra{-},
\label{eq:pvm_kraus}
\end{equation}
where $\ket{\pm} = (\ket{0} \pm \ket{1})/\sqrt{2}$. This means that we can interpret this particular $\Phi$ as a dephasing channel. The POVM elements associated with Eq.~\eqref{eq:pvm_kraus} are $E_\pm = K_\pm$, as one would expect from Eq.~\eqref{eq:state_update}, which describes the target state update rule of a projective $\sigma_x^{(\textsc{t})}$ measurement. The channel describes how arbitrary input states are mapped after the idealized measurement into the state that is the correct probabilistic mixture of the pure states associated that describe the possible measurement outcomes.  

%{\red Notice that, while the interaction described above is projective in nature (i.e., it eliminates off-diagonal terms in the measured basis), the resulting POVM elements are not projections, and thus do not form a PVM. It is worth emphasizing that this is due to the enlarged Heisenberg cut: When treating detectors quantum mechanically, a measurement does not collapse a state to a single outcome, but to a probabilistic ensemble of all outcomes, as mentioend above after Eq.~\eqref{eq:POVM_update}. Thus, it is possible for all POVM elements of a non-trivial measurement to be proportional to $I^{(\textsc{t})}$ in this setting (even though the measurement operators $\{ K_i\}$ are generally non-trivial).}

%Notice that, due to the expanded Heisenberg cut, they do not represent a PVM. {\color{red} \bf It is worth emphasizing again that this is due to the enlarged Heisenberg cut: When treating detectors quantum mechanically, a measurement does not collapse a state to a single outcome, but to a probabilistic ensemble of all outcomes, as mentioend above after Eq.~\eqref{eq:POVM_update}.} Thus, it is possible for all POVM elements of a non-trivial measurement to be proportional to $I^{(\textsc{t})}$ in this setting. \rm

\subsection{A realistic example}

Let us now consider an example of a particular Zeno-like fixed point resulting from finite frequency measurements with a finite coupling strength. For the parameter values
\begin{equation}
\label{eq:params}
\Delta t_f = 15.13 \, \omega^{-1}
\quad
\Delta t_m = 14.96 \, \omega^{-1}
\quad
\frac{g}{\omega} = 0.865,
\end{equation}
the channel $\Phi$ has the eigenvectors
\begin{align}
V_0' &= \begin{pmatrix}
0.5 & 0\\
0 & 0.5
\end{pmatrix}_{[\mathcal{Z}]}  \quad
V_1' = \begin{pmatrix}
1.0 & 0 \\
0 & -1.0
\end{pmatrix}_{[\mathcal{Z}]}\nonumber\\
V_2' &= 
\begin{pmatrix}
0 & 0.42 - 0.27\ii\\
0.42 + 0.27\ii & 0
\end{pmatrix}_{[\mathcal{Z}]} \label{eq:realistic_vecs}\\
V_3' &= \begin{pmatrix}
0 & 0.27 + 0.42 \ii\\
0.24 - 0.42 \ii& 0
\end{pmatrix}_{[\mathcal{Z}]},
\nonumber
\end{align}
and the eigenvalues
\begin{align}\label{eq:eigsex}
\lambda_0' &= 1.0 &
\lambda_1' &= 0.73 \\
\lambda_2' &= 1.0 &
\lambda_3' &= 0.73. \nonumber
\end{align}
Thus, we have that $\Phi \big(V_0' + \alpha V_2'\big) = V_0' + \alpha V_2'$ for $\alpha \in [-1,1]$. In particular, the pure states $V_0' \pm V_2'$ are perfectly preserved (to numerical precision) by the Zeno-like effect originating from finite-frequency measurements.

A set of Kraus operators for $\Phi$ in this example is
\begin{align}
K_\pm' = (0.36+0.55\ii) \, I^{(\textsc{t})} \pm 0.22 \, \sigma_x^{(\textsc{t})} \pm 0.14 \, \sigma_y^{(\textsc{t})} \label{eq:kraus_example}
\end{align}

One can readily check that the results in Eqs.~\eqref{eq:realistic_vecs}--\eqref{eq:kraus_example} are stable under small perturbations of the parameters in Eq.~\eqref{eq:params}. Observe that $\Phi$ represents a non-trivial---albeit non-projective---measurement in this case. The POVM elements arising from Eq.~\eqref{eq:kraus_example} are
\begin{align}
E_\pm' = \frac{1}{2} \big( I^{(\textsc{t})} \pm \boldsymbol{n} \cdot \boldsymbol{\sigma}^{(\textsc{t})} \big),
\end{align}

where $\boldsymbol{n} = (0.32, 0.20, 0)$.

%%%%%%%%%%%%%%%%%%%%%%%%%%%%%%%%%%%%%%%%%%
\section{Conclusions and Outlook}
We conclude that the Zeno effect is a special instance of a more general Zeno phenomenon that allows one to effectively halt the evolution of a target system through measurements, even if they are imperfect and repeated only at a finite frequency. Also, in contrast to the conventional Zeno effect, in this more general scheme the preserved states need not be eigenstates of the interaction Hamiltonian and instead can be placed anywhere in the Bloch sphere by suitably choosing measurement parameters such as the interaction time and coupling strength. 

Generally, imperfect measurements performed at a finite frequency do not freeze the target system, of course. However, we found that any one of a wide range of states can be arranged to be a Zeno-like fixed point for the target system by suitably choosing the measurement parameter values, as shown in Fig.~\ref{fig:1b}. The conventional Zeno fixed point is merely that special case of these Zeno-like fixed points where the measurement parameters approach perfect measurements at infinite frequency.  

Further, when the measurement parameters are not specially chosen, and when there is, therefore, no Zeno-like fixed point, then the target system's state will eventually approach a state that is a universally attractive fixed point. We showed that the existence of this fixed point is guaranteed by applying Brouwer's theorem. One might have expected this `Brouwer fixed point' to be generally maximally mixed. Surprisingly, however, as Fig.~\ref{fig:1c} shows, the Brouwer fixed points can be chosen all over the Bloch sphere, by suitably choosing the initial state of the detectors and the measurement parameters. In experiments, whether Zeno-like or Brouwer fixed points are more readily implementable depends on how well and in what range the measurement parameters and the initial state of the detectors can be prepared.  

The Zeno-like fixed points are similar in nature to the conventional Zeno fixed point because the latter is a special case. There is a different intuition for the Brouwer fixed points: 
Consider the case where the detectors are prepared in the ground state, i.e., cold. On one hand, the detectors will therefore generally take away heat from the target. On the other hand, the suddenness of the target-detector interactions tends to heat the target. After transients, the resulting fixed point state, which is of course generally not a thermodynamical equilibrium state, is the Brouwer fixed point state.      

While in this  paper we mostly focused on qubits, we expect Zeno-like and Brouwer fixed points to be present much more generally. Consider, for example, the setup of entanglement farming from an optical cavity field \cite{Martinez:2013}. There, identically-prepared unentangled pairs of atoms are successively sent through an optical cavity. The successive pairs of atoms were observed to drive the cavity field towards a certain entangling state. While no connection to the Zeno effect was made in \cite{Martinez:2013}, we can now see that this entangling state is likely an instance of a Zeno-like fixed point, with each pair of atoms playing the role of a detector and the cavity field being the target system. In addition, the existence of Brouwer-type universally attractive fixed points for systems with higher dimensional, or even infinite dimensional Hilbert spaces can be shown, as above, using Brouwer \cite{Brouwer} and Schauder's theorems \cite{Schauder}, respectively. 

Finally, we note that it is also possible for repeated measurements to enhance---rather than freeze---evolution; a phenomenon known as the anti-Zeno effect \cite{Ai:2013, Fischer:2001, Ruseckas:2006, Kofman:2000}. It will be interesting to study also this phenomenon with the approach that we pursued here.

\section{Acknowledgments}
AK, EMM and DL acknowledge support from the NSERC Discovery, NSERC Banting, and NSERC PGSM programs respectively.

\appendix

\section{Eigenpairs of the channel $\Phi$}

In the general case of finite frequency measurements, certain eigenpairs of $\Phi$ are highly complicated functions of the measurement parameters ($\Delta t_f$, $\Delta t_m$, $\omega$, $g$). In the main text, we gave explicit expressions for two particularly simple eigenvalues, and restricted ourselves to a more qualitative discussion of the other eigenvalues and eigenvectors.

Here, we provide a matrix representation for $\Phi$, from which the eigenpairs can easily be extracted using any computer algebra system. The rationale is that the expressions for certain eigenpairs have on the order of 1000 terms, and are thus too lengthy to be useful when written out explicitly. In particular, we define a $4 \times 4$ matrix $M^\Phi$, such that if $A = \Phi(B)$, where $A, B \in \mathcal{L}( \mathcal{H}^{(\textsc{t})})$, then
\begin{equation}
\text{vec}(A) = M^\Phi \text{vec}(B),
\end{equation}
where the vectorization operation is defined as
\begin{equation}
\text{vec}
\begin{pmatrix}
a_{1,1} & a_{1,2} \\ a_{2,1} & a_{2,2}
\end{pmatrix}
\equiv
\begin{pmatrix}
a_{1,1}\\ a_{2,1}\\a_{1,2}\\a_{2,2}
\end{pmatrix}.
\end{equation}

The eigenvalues of $\Phi$ are the same as those of $M^\Phi$. The eigenvectors of $\Phi$ can be determined by inverting the vectorization operation on those of the $M^\Phi$. The nonzero elements of $M^\Phi$ are as follows:

\begin{widetext}
\begin{equation}
M^\Phi_{11} =  \frac{
2\left[ \left( g^3\Omega + g^4 + 2g^2\omega^2\right) \cos^2\left( \Omega \Delta t_m\right) + 4\Omega g \omega^2 + 4g^2\omega^2 + 8\omega^4\right]}{
\Omega^2 \left( g + \Omega \right)^2
},
\end{equation}
\begin{equation}
M^\Phi_{1,4} = \frac{
\left(\Omega g + g^2 + 4 \omega^2\right)^2 \sin^2 \left(\Omega \Delta t_m \right)
}{
\Omega^2 \left( g + \Omega \right)^2
},
\end{equation}
\begin{equation}
M^\Phi_{2,2} = \overline{M^\Phi_{3,3}} = 
\frac{ 
\cos(g \Delta t_m) (\Omega g + \Omega^2) e^{2i\omega \Delta t_f}
\left[ (\Omega g + g^2 + 4 \omega^2) \cos(\Omega \Delta t_m) + 2i( \omega \Omega + 2 \omega g) \sin(\Omega \Delta t_m)
\right]
}{
\Omega^2 \left( g + \Omega \right)^2
},
\end{equation}
\begin{equation}
M^\Phi_{2,3} = \overline{M^\Phi_{3,2}} = \frac{
e^{-2i\omega \Delta t_f} g \sin(\Omega \Delta t_m) \sin(g \Delta t_m) (\Omega g + g^2 + 4 \omega^2)
}{
\Omega^2 \left( g + \Omega \right)^2
},
\end{equation}
\begin{equation}
M^\Phi_{4,1} = \frac{
g^2 \sin^2 (\Omega \Delta t_m)
}{
\Omega^2
},
\end{equation}
\begin{equation}
M^\Phi_{4,4} = \frac{
\left(\Omega g + g^2 + 4 \omega^2\right)^2 \cos^2 \left(\Omega \Delta t_m \right)
}{
\Omega^2 \left( g + \Omega \right)^2,
}
\end{equation}
where $\Omega = \sqrt{g^2 + 4\omega^2}$.
\end{widetext}

\bibliography{References}

\begin{thebibliography}{31}%
\makeatletter
\providecommand \@ifxundefined [1]{%
 \@ifx{#1\undefined}
}%
\providecommand \@ifnum [1]{%
 \ifnum #1\expandafter \@firstoftwo
 \else \expandafter \@secondoftwo
 \fi
}%
\providecommand \@ifx [1]{%
 \ifx #1\expandafter \@firstoftwo
 \else \expandafter \@secondoftwo
 \fi
}%
\providecommand \natexlab [1]{#1}%
\providecommand \enquote  [1]{``#1''}%
\providecommand \bibnamefont  [1]{#1}%
\providecommand \bibfnamefont [1]{#1}%
\providecommand \citenamefont [1]{#1}%
\providecommand \href@noop [0]{\@secondoftwo}%
\providecommand \href [0]{\begingroup \@sanitize@url \@href}%
\providecommand \@href[1]{\@@startlink{#1}\@@href}%
\providecommand \@@href[1]{\endgroup#1\@@endlink}%
\providecommand \@sanitize@url [0]{\catcode `\\12\catcode `\$12\catcode
  `\&12\catcode `\#12\catcode `\^12\catcode `\_12\catcode `\%12\relax}%
\providecommand \@@startlink[1]{}%
\providecommand \@@endlink[0]{}%
\providecommand \url  [0]{\begingroup\@sanitize@url \@url }%
\providecommand \@url [1]{\endgroup\@href {#1}{\urlprefix }}%
\providecommand \urlprefix  [0]{URL }%
\providecommand \Eprint [0]{\href }%
\providecommand \doibase [0]{http://dx.doi.org/}%
\providecommand \selectlanguage [0]{\@gobble}%
\providecommand \bibinfo  [0]{\@secondoftwo}%
\providecommand \bibfield  [0]{\@secondoftwo}%
\providecommand \translation [1]{[#1]}%
\providecommand \BibitemOpen [0]{}%
\providecommand \bibitemStop [0]{}%
\providecommand \bibitemNoStop [0]{.\EOS\space}%
\providecommand \EOS [0]{\spacefactor3000\relax}%
\providecommand \BibitemShut  [1]{\csname bibitem#1\endcsname}%
\let\auto@bib@innerbib\@empty
%</preamble>
\bibitem [{\citenamefont {Erez}\ \emph {et~al.}(2004)\citenamefont {Erez},
  \citenamefont {Aharonov}, \citenamefont {Reznik},\ and\ \citenamefont
  {Vaidman}}]{Erez:2004}%
  \BibitemOpen
  \bibfield  {author} {\bibinfo {author} {\bibfnamefont {N.}~\bibnamefont
  {Erez}}, \bibinfo {author} {\bibfnamefont {Y.}~\bibnamefont {Aharonov}},
  \bibinfo {author} {\bibfnamefont {B.}~\bibnamefont {Reznik}}, \ and\ \bibinfo
  {author} {\bibfnamefont {L.}~\bibnamefont {Vaidman}},\ }\href {\doibase
  10.1103/PhysRevA.69.062315} {\bibfield  {journal} {\bibinfo  {journal} {Phys.
  Rev. A}\ }\textbf {\bibinfo {volume} {69}},\ \bibinfo {pages} {062315}
  (\bibinfo {year} {2004})}\BibitemShut {NoStop}%
\bibitem [{\citenamefont {Paz-Silva}\ \emph {et~al.}(2012)\citenamefont
  {Paz-Silva}, \citenamefont {Rezakhani}, \citenamefont {Dominy},\ and\
  \citenamefont {Lidar}}]{Silva:2012}%
  \BibitemOpen
  \bibfield  {author} {\bibinfo {author} {\bibfnamefont {G.~A.}\ \bibnamefont
  {Paz-Silva}}, \bibinfo {author} {\bibfnamefont {A.~T.}\ \bibnamefont
  {Rezakhani}}, \bibinfo {author} {\bibfnamefont {J.~M.}\ \bibnamefont
  {Dominy}}, \ and\ \bibinfo {author} {\bibfnamefont {D.~A.}\ \bibnamefont
  {Lidar}},\ }\href {\doibase 10.1103/PhysRevLett.108.080501} {\bibfield
  {journal} {\bibinfo  {journal} {Phys. Rev. Lett.}\ }\textbf {\bibinfo
  {volume} {108}},\ \bibinfo {pages} {080501} (\bibinfo {year}
  {2012})}\BibitemShut {NoStop}%
\bibitem [{\citenamefont {Dominy}\ \emph {et~al.}(2013)\citenamefont {Dominy},
  \citenamefont {Paz-Silva}, \citenamefont {Rezakhani},\ and\ \citenamefont
  {Lidar}}]{Dominy:2013}%
  \BibitemOpen
  \bibfield  {author} {\bibinfo {author} {\bibfnamefont {J.~M.}\ \bibnamefont
  {Dominy}}, \bibinfo {author} {\bibfnamefont {G.~A.}\ \bibnamefont
  {Paz-Silva}}, \bibinfo {author} {\bibfnamefont {A.~T.}\ \bibnamefont
  {Rezakhani}}, \ and\ \bibinfo {author} {\bibfnamefont {D.~A.}\ \bibnamefont
  {Lidar}},\ }\href {http://stacks.iop.org/1751-8121/46/i=7/a=075306}
  {\bibfield  {journal} {\bibinfo  {journal} {J. Phys. A}\ }\textbf {\bibinfo
  {volume} {46}},\ \bibinfo {pages} {075306} (\bibinfo {year}
  {2013})}\BibitemShut {NoStop}%
\bibitem [{\citenamefont {Beige}\ \emph {et~al.}(2000)\citenamefont {Beige},
  \citenamefont {Braun}, \citenamefont {Tregenna},\ and\ \citenamefont
  {Knight}}]{Beige:2000}%
  \BibitemOpen
  \bibfield  {author} {\bibinfo {author} {\bibfnamefont {A.}~\bibnamefont
  {Beige}}, \bibinfo {author} {\bibfnamefont {D.}~\bibnamefont {Braun}},
  \bibinfo {author} {\bibfnamefont {B.}~\bibnamefont {Tregenna}}, \ and\
  \bibinfo {author} {\bibfnamefont {P.~L.}\ \bibnamefont {Knight}},\ }\href
  {\doibase 10.1103/PhysRevLett.85.1762} {\bibfield  {journal} {\bibinfo
  {journal} {Phys. Rev. Lett.}\ }\textbf {\bibinfo {volume} {85}},\ \bibinfo
  {pages} {1762} (\bibinfo {year} {2000})}\BibitemShut {NoStop}%
\bibitem [{\citenamefont {Nakazato}\ \emph {et~al.}(2003)\citenamefont
  {Nakazato}, \citenamefont {Takazawa},\ and\ \citenamefont
  {Yuasa}}]{Nakazato:2003}%
  \BibitemOpen
  \bibfield  {author} {\bibinfo {author} {\bibfnamefont {H.}~\bibnamefont
  {Nakazato}}, \bibinfo {author} {\bibfnamefont {T.}~\bibnamefont {Takazawa}},
  \ and\ \bibinfo {author} {\bibfnamefont {K.}~\bibnamefont {Yuasa}},\ }\href
  {\doibase 10.1103/PhysRevLett.90.060401} {\bibfield  {journal} {\bibinfo
  {journal} {Phys. Rev. Lett.}\ }\textbf {\bibinfo {volume} {90}},\ \bibinfo
  {pages} {060401} (\bibinfo {year} {2003})}\BibitemShut {NoStop}%
\bibitem [{\citenamefont {Nakazato}\ \emph {et~al.}(2004)\citenamefont
  {Nakazato}, \citenamefont {Unoki},\ and\ \citenamefont
  {Yuasa}}]{Nakazato:2004}%
  \BibitemOpen
  \bibfield  {author} {\bibinfo {author} {\bibfnamefont {H.}~\bibnamefont
  {Nakazato}}, \bibinfo {author} {\bibfnamefont {M.}~\bibnamefont {Unoki}}, \
  and\ \bibinfo {author} {\bibfnamefont {K.}~\bibnamefont {Yuasa}},\ }\href
  {\doibase 10.1103/PhysRevA.70.012303} {\bibfield  {journal} {\bibinfo
  {journal} {Phys. Rev. A}\ }\textbf {\bibinfo {volume} {70}},\ \bibinfo
  {pages} {012303} (\bibinfo {year} {2004})}\BibitemShut {NoStop}%
\bibitem [{\citenamefont {Wang}\ \emph {et~al.}(2008)\citenamefont {Wang},
  \citenamefont {You},\ and\ \citenamefont {Nori}}]{Wang:2008}%
  \BibitemOpen
  \bibfield  {author} {\bibinfo {author} {\bibfnamefont {X.-B.}\ \bibnamefont
  {Wang}}, \bibinfo {author} {\bibfnamefont {J.~Q.}\ \bibnamefont {You}}, \
  and\ \bibinfo {author} {\bibfnamefont {F.}~\bibnamefont {Nori}},\ }\href
  {\doibase 10.1103/PhysRevA.77.062339} {\bibfield  {journal} {\bibinfo
  {journal} {Phys. Rev. A}\ }\textbf {\bibinfo {volume} {77}},\ \bibinfo
  {pages} {062339} (\bibinfo {year} {2008})}\BibitemShut {NoStop}%
\bibitem [{\citenamefont {Franson}\ \emph {et~al.}(2004)\citenamefont
  {Franson}, \citenamefont {Jacobs},\ and\ \citenamefont
  {Pittman}}]{Franson:2004}%
  \BibitemOpen
  \bibfield  {author} {\bibinfo {author} {\bibfnamefont {J.~D.}\ \bibnamefont
  {Franson}}, \bibinfo {author} {\bibfnamefont {B.~C.}\ \bibnamefont {Jacobs}},
  \ and\ \bibinfo {author} {\bibfnamefont {T.~B.}\ \bibnamefont {Pittman}},\
  }\href {\doibase 10.1103/PhysRevA.70.062302} {\bibfield  {journal} {\bibinfo
  {journal} {Phys. Rev. A}\ }\textbf {\bibinfo {volume} {70}},\ \bibinfo
  {pages} {062302} (\bibinfo {year} {2004})}\BibitemShut {NoStop}%
\bibitem [{\citenamefont {Maniscalco}\ \emph {et~al.}(2008)\citenamefont
  {Maniscalco}, \citenamefont {Francica}, \citenamefont {Zaffino},
  \citenamefont {Lo~Gullo},\ and\ \citenamefont {Plastina}}]{Maniscalco:2008}%
  \BibitemOpen
  \bibfield  {author} {\bibinfo {author} {\bibfnamefont {S.}~\bibnamefont
  {Maniscalco}}, \bibinfo {author} {\bibfnamefont {F.}~\bibnamefont
  {Francica}}, \bibinfo {author} {\bibfnamefont {R.~L.}\ \bibnamefont
  {Zaffino}}, \bibinfo {author} {\bibfnamefont {N.}~\bibnamefont {Lo~Gullo}}, \
  and\ \bibinfo {author} {\bibfnamefont {F.}~\bibnamefont {Plastina}},\ }\href
  {\doibase 10.1103/PhysRevLett.100.090503} {\bibfield  {journal} {\bibinfo
  {journal} {Phys. Rev. Lett.}\ }\textbf {\bibinfo {volume} {100}},\ \bibinfo
  {pages} {090503} (\bibinfo {year} {2008})}\BibitemShut {NoStop}%
\bibitem [{\citenamefont {Bernu}\ \emph {et~al.}(2008)\citenamefont {Bernu},
  \citenamefont {Del\'eglise}, \citenamefont {Sayrin}, \citenamefont {Kuhr},
  \citenamefont {Dotsenko}, \citenamefont {Brune}, \citenamefont {Raimond},\
  and\ \citenamefont {Haroche}}]{Bernu:2008}%
  \BibitemOpen
  \bibfield  {author} {\bibinfo {author} {\bibfnamefont {J.}~\bibnamefont
  {Bernu}}, \bibinfo {author} {\bibfnamefont {S.}~\bibnamefont {Del\'eglise}},
  \bibinfo {author} {\bibfnamefont {C.}~\bibnamefont {Sayrin}}, \bibinfo
  {author} {\bibfnamefont {S.}~\bibnamefont {Kuhr}}, \bibinfo {author}
  {\bibfnamefont {I.}~\bibnamefont {Dotsenko}}, \bibinfo {author}
  {\bibfnamefont {M.}~\bibnamefont {Brune}}, \bibinfo {author} {\bibfnamefont
  {J.~M.}\ \bibnamefont {Raimond}}, \ and\ \bibinfo {author} {\bibfnamefont
  {S.}~\bibnamefont {Haroche}},\ }\href {\doibase
  10.1103/PhysRevLett.101.180402} {\bibfield  {journal} {\bibinfo  {journal}
  {Phys. Rev. Lett.}\ }\textbf {\bibinfo {volume} {101}},\ \bibinfo {pages}
  {180402} (\bibinfo {year} {2008})}\BibitemShut {NoStop}%
\bibitem [{\citenamefont {Raimond}\ \emph {et~al.}(2012)\citenamefont
  {Raimond}, \citenamefont {Facchi}, \citenamefont {Peaudecerf}, \citenamefont
  {Pascazio}, \citenamefont {Sayrin}, \citenamefont {Dotsenko}, \citenamefont
  {Gleyzes}, \citenamefont {Brune},\ and\ \citenamefont
  {Haroche}}]{Raimond:2012}%
  \BibitemOpen
  \bibfield  {author} {\bibinfo {author} {\bibfnamefont {J.~M.}\ \bibnamefont
  {Raimond}}, \bibinfo {author} {\bibfnamefont {P.}~\bibnamefont {Facchi}},
  \bibinfo {author} {\bibfnamefont {B.}~\bibnamefont {Peaudecerf}}, \bibinfo
  {author} {\bibfnamefont {S.}~\bibnamefont {Pascazio}}, \bibinfo {author}
  {\bibfnamefont {C.}~\bibnamefont {Sayrin}}, \bibinfo {author} {\bibfnamefont
  {I.}~\bibnamefont {Dotsenko}}, \bibinfo {author} {\bibfnamefont
  {S.}~\bibnamefont {Gleyzes}}, \bibinfo {author} {\bibfnamefont
  {M.}~\bibnamefont {Brune}}, \ and\ \bibinfo {author} {\bibfnamefont
  {S.}~\bibnamefont {Haroche}},\ }\href {\doibase 10.1103/PhysRevA.86.032120}
  {\bibfield  {journal} {\bibinfo  {journal} {Phys. Rev. A}\ }\textbf {\bibinfo
  {volume} {86}},\ \bibinfo {pages} {032120} (\bibinfo {year}
  {2012})}\BibitemShut {NoStop}%
\bibitem [{\citenamefont {Itano}\ \emph {et~al.}(1990)\citenamefont {Itano},
  \citenamefont {Heinzen}, \citenamefont {Bollinger},\ and\ \citenamefont
  {Wineland}}]{Itano:1990}%
  \BibitemOpen
  \bibfield  {author} {\bibinfo {author} {\bibfnamefont {W.~M.}\ \bibnamefont
  {Itano}}, \bibinfo {author} {\bibfnamefont {D.~J.}\ \bibnamefont {Heinzen}},
  \bibinfo {author} {\bibfnamefont {J.~J.}\ \bibnamefont {Bollinger}}, \ and\
  \bibinfo {author} {\bibfnamefont {D.~J.}\ \bibnamefont {Wineland}},\ }\href
  {\doibase 10.1103/PhysRevA.41.2295} {\bibfield  {journal} {\bibinfo
  {journal} {Phys. Rev. A}\ }\textbf {\bibinfo {volume} {41}},\ \bibinfo
  {pages} {2295} (\bibinfo {year} {1990})}\BibitemShut {NoStop}%
\bibitem [{\citenamefont {Fischer}\ \emph {et~al.}(2001)\citenamefont
  {Fischer}, \citenamefont {Guti\'errez-Medina},\ and\ \citenamefont
  {Raizen}}]{Fischer:2001}%
  \BibitemOpen
  \bibfield  {author} {\bibinfo {author} {\bibfnamefont {M.~C.}\ \bibnamefont
  {Fischer}}, \bibinfo {author} {\bibfnamefont {B.}~\bibnamefont
  {Guti\'errez-Medina}}, \ and\ \bibinfo {author} {\bibfnamefont {M.~G.}\
  \bibnamefont {Raizen}},\ }\href {\doibase 10.1103/PhysRevLett.87.040402}
  {\bibfield  {journal} {\bibinfo  {journal} {Phys. Rev. Lett.}\ }\textbf
  {\bibinfo {volume} {87}},\ \bibinfo {pages} {040402} (\bibinfo {year}
  {2001})}\BibitemShut {NoStop}%
\bibitem [{\citenamefont {Xiao}\ and\ \citenamefont {Jones}(2006)}]{Xiao:2006}%
  \BibitemOpen
  \bibfield  {author} {\bibinfo {author} {\bibfnamefont {L.}~\bibnamefont
  {Xiao}}\ and\ \bibinfo {author} {\bibfnamefont {J.~A.}\ \bibnamefont
  {Jones}},\ }\href {\doibase http://dx.doi.org/10.1016/j.physleta.2006.06.086}
  {\bibfield  {journal} {\bibinfo  {journal} {Phys. Lett. A}\ }\textbf
  {\bibinfo {volume} {359}},\ \bibinfo {pages} {424 } (\bibinfo {year}
  {2006})}\BibitemShut {NoStop}%
\bibitem [{\citenamefont {Zheng}\ \emph {et~al.}(2013)\citenamefont {Zheng},
  \citenamefont {Xu}, \citenamefont {Peng}, \citenamefont {Zhou}, \citenamefont
  {Du},\ and\ \citenamefont {Sun}}]{Zheng:2013}%
  \BibitemOpen
  \bibfield  {author} {\bibinfo {author} {\bibfnamefont {W.}~\bibnamefont
  {Zheng}}, \bibinfo {author} {\bibfnamefont {D.~Z.}\ \bibnamefont {Xu}},
  \bibinfo {author} {\bibfnamefont {X.}~\bibnamefont {Peng}}, \bibinfo {author}
  {\bibfnamefont {X.}~\bibnamefont {Zhou}}, \bibinfo {author} {\bibfnamefont
  {J.}~\bibnamefont {Du}}, \ and\ \bibinfo {author} {\bibfnamefont {C.~P.}\
  \bibnamefont {Sun}},\ }\href {\doibase 10.1103/PhysRevA.87.032112} {\bibfield
   {journal} {\bibinfo  {journal} {Phys. Rev. A}\ }\textbf {\bibinfo {volume}
  {87}},\ \bibinfo {pages} {032112} (\bibinfo {year} {2013})}\BibitemShut
  {NoStop}%
\bibitem [{\citenamefont {Petrosky}\ \emph {et~al.}(1990)\citenamefont
  {Petrosky}, \citenamefont {Tasaki},\ and\ \citenamefont
  {Prigogine}}]{Petrosky:1990}%
  \BibitemOpen
  \bibfield  {author} {\bibinfo {author} {\bibfnamefont {T.}~\bibnamefont
  {Petrosky}}, \bibinfo {author} {\bibfnamefont {S.}~\bibnamefont {Tasaki}}, \
  and\ \bibinfo {author} {\bibfnamefont {I.}~\bibnamefont {Prigogine}},\ }\href
  {\doibase http://dx.doi.org/10.1016/0375-9601(90)90173-L} {\bibfield
  {journal} {\bibinfo  {journal} {Phys. Lett. A}\ }\textbf {\bibinfo {volume}
  {151}},\ \bibinfo {pages} {109 } (\bibinfo {year} {1990})}\BibitemShut
  {NoStop}%
\bibitem [{\citenamefont {Peres}\ and\ \citenamefont {Ron}(1990)}]{Peres:1990}%
  \BibitemOpen
  \bibfield  {author} {\bibinfo {author} {\bibfnamefont {A.}~\bibnamefont
  {Peres}}\ and\ \bibinfo {author} {\bibfnamefont {A.}~\bibnamefont {Ron}},\
  }\href {\doibase 10.1103/PhysRevA.42.5720} {\bibfield  {journal} {\bibinfo
  {journal} {Phys. Rev. A}\ }\textbf {\bibinfo {volume} {42}},\ \bibinfo
  {pages} {5720} (\bibinfo {year} {1990})}\BibitemShut {NoStop}%
\bibitem [{\citenamefont {Pascazio}\ and\ \citenamefont
  {Namiki}(1994)}]{Pascazio:1994}%
  \BibitemOpen
  \bibfield  {author} {\bibinfo {author} {\bibfnamefont {S.}~\bibnamefont
  {Pascazio}}\ and\ \bibinfo {author} {\bibfnamefont {M.}~\bibnamefont
  {Namiki}},\ }\href {\doibase 10.1103/PhysRevA.50.4582} {\bibfield  {journal}
  {\bibinfo  {journal} {Phys. Rev. A}\ }\textbf {\bibinfo {volume} {50}},\
  \bibinfo {pages} {4582} (\bibinfo {year} {1994})}\BibitemShut {NoStop}%
\bibitem [{\citenamefont {Ruseckas}\ and\ \citenamefont
  {Kaulakys}(2001)}]{Ruseckas:2001}%
  \BibitemOpen
  \bibfield  {author} {\bibinfo {author} {\bibfnamefont {J.}~\bibnamefont
  {Ruseckas}}\ and\ \bibinfo {author} {\bibfnamefont {B.}~\bibnamefont
  {Kaulakys}},\ }\href {\doibase 10.1103/PhysRevA.63.062103} {\bibfield
  {journal} {\bibinfo  {journal} {Phys. Rev. A}\ }\textbf {\bibinfo {volume}
  {63}},\ \bibinfo {pages} {062103} (\bibinfo {year} {2001})}\BibitemShut
  {NoStop}%
\bibitem [{\citenamefont {Facchi}\ and\ \citenamefont
  {Pascazio}(2002)}]{Facchi:2002}%
  \BibitemOpen
  \bibfield  {author} {\bibinfo {author} {\bibfnamefont {P.}~\bibnamefont
  {Facchi}}\ and\ \bibinfo {author} {\bibfnamefont {S.}~\bibnamefont
  {Pascazio}},\ }\href {\doibase 10.1103/PhysRevLett.89.080401} {\bibfield
  {journal} {\bibinfo  {journal} {Phys. Rev. Lett.}\ }\textbf {\bibinfo
  {volume} {89}},\ \bibinfo {pages} {080401} (\bibinfo {year}
  {2002})}\BibitemShut {NoStop}%
\bibitem [{\citenamefont {Ai}\ \emph {et~al.}()\citenamefont {Ai},
  \citenamefont {Xu}, \citenamefont {Yi}, \citenamefont {Kofman}, \citenamefont
  {Sun},\ and\ \citenamefont {Nori}}]{Ai:2013}%
  \BibitemOpen
  \bibfield  {author} {\bibinfo {author} {\bibfnamefont {Q.}~\bibnamefont
  {Ai}}, \bibinfo {author} {\bibfnamefont {D.}~\bibnamefont {Xu}}, \bibinfo
  {author} {\bibfnamefont {S.}~\bibnamefont {Yi}}, \bibinfo {author}
  {\bibfnamefont {A.}~\bibnamefont {Kofman}}, \bibinfo {author} {\bibfnamefont
  {C.}~\bibnamefont {Sun}}, \ and\ \bibinfo {author} {\bibfnamefont
  {F.}~\bibnamefont {Nori}},\ }\href@noop {} {\bibfield  {journal} {\bibinfo
  {journal} {Sci. Rep.}\ }\textbf {\bibinfo {volume} {3}},\ \bibinfo {pages}
  {1752}}\BibitemShut {NoStop}%
\bibitem [{\citenamefont {Brouwer}(1911)}]{Brouwer}%
  \BibitemOpen
  \bibfield  {author} {\bibinfo {author} {\bibfnamefont {L.}~\bibnamefont
  {Brouwer}},\ }\href {\doibase 10.1007/BF01456931} {\bibfield  {journal}
  {\bibinfo  {journal} {Math. Ann.}\ }\textbf {\bibinfo {volume} {71}},\
  \bibinfo {pages} {97} (\bibinfo {year} {1911})}\BibitemShut {NoStop}%
\bibitem [{\citenamefont {Scully}\ and\ \citenamefont
  {Zubairy}(1997)}]{Scully:1997}%
  \BibitemOpen
  \bibfield  {author} {\bibinfo {author} {\bibfnamefont {M.}~\bibnamefont
  {Scully}}\ and\ \bibinfo {author} {\bibfnamefont {S.}~\bibnamefont
  {Zubairy}},\ }\href {http://books.google.ca/books?id=20ISsQCKKmQC} {\emph
  {\bibinfo {title} {Quantum Optics}}}\ (\bibinfo  {publisher} {Cambridge
  University Press, Cambridge, England},\ \bibinfo {year} {1997})\BibitemShut
  {NoStop}%
\bibitem [{\citenamefont {De~Groot}\ \emph {et~al.}(2010)\citenamefont
  {De~Groot}, \citenamefont {Lisenfeld}, \citenamefont {Schouten},
  \citenamefont {Ashhab}, \citenamefont {Lupa{\c{s}}cu}, \citenamefont
  {Harmans},\ and\ \citenamefont {Mooij}}]{degroot:2010}%
  \BibitemOpen
  \bibfield  {author} {\bibinfo {author} {\bibfnamefont {P.}~\bibnamefont
  {De~Groot}}, \bibinfo {author} {\bibfnamefont {J.}~\bibnamefont {Lisenfeld}},
  \bibinfo {author} {\bibfnamefont {R.}~\bibnamefont {Schouten}}, \bibinfo
  {author} {\bibfnamefont {S.}~\bibnamefont {Ashhab}}, \bibinfo {author}
  {\bibfnamefont {A.}~\bibnamefont {Lupa{\c{s}}cu}}, \bibinfo {author}
  {\bibfnamefont {C.}~\bibnamefont {Harmans}}, \ and\ \bibinfo {author}
  {\bibfnamefont {J.}~\bibnamefont {Mooij}},\ }\href@noop {} {\bibfield
  {journal} {\bibinfo  {journal} {Nat. Phys.}\ }\textbf {\bibinfo {volume}
  {6}},\ \bibinfo {pages} {763} (\bibinfo {year} {2010})}\BibitemShut {NoStop}%
\bibitem [{\citenamefont {Majer}\ \emph {et~al.}(2007)\citenamefont {Majer},
  \citenamefont {Chow}, \citenamefont {Gambetta}, \citenamefont {Koch},
  \citenamefont {Johnson}, \citenamefont {Schreier}, \citenamefont {Frunzio},
  \citenamefont {Schuster}, \citenamefont {Houck}, \citenamefont {Wallraff}
  \emph {et~al.}}]{Majer:2007}%
  \BibitemOpen
  \bibfield  {author} {\bibinfo {author} {\bibfnamefont {J.}~\bibnamefont
  {Majer}}, \bibinfo {author} {\bibfnamefont {J.}~\bibnamefont {Chow}},
  \bibinfo {author} {\bibfnamefont {J.}~\bibnamefont {Gambetta}}, \bibinfo
  {author} {\bibfnamefont {J.}~\bibnamefont {Koch}}, \bibinfo {author}
  {\bibfnamefont {B.}~\bibnamefont {Johnson}}, \bibinfo {author} {\bibfnamefont
  {J.}~\bibnamefont {Schreier}}, \bibinfo {author} {\bibfnamefont
  {L.}~\bibnamefont {Frunzio}}, \bibinfo {author} {\bibfnamefont
  {D.}~\bibnamefont {Schuster}}, \bibinfo {author} {\bibfnamefont
  {A.}~\bibnamefont {Houck}}, \bibinfo {author} {\bibfnamefont
  {A.}~\bibnamefont {Wallraff}},  \emph {et~al.},\ }\href@noop {} {\bibfield
  {journal} {\bibinfo  {journal} {Nature (London)}\ }\textbf {\bibinfo {volume}
  {449}},\ \bibinfo {pages} {443} (\bibinfo {year} {2007})}\BibitemShut
  {NoStop}%
\bibitem [{\citenamefont {Ficek}\ and\ \citenamefont
  {Wahiddin}(2014)}]{Ficek:2004}%
  \BibitemOpen
  \bibfield  {author} {\bibinfo {author} {\bibfnamefont {Z.}~\bibnamefont
  {Ficek}}\ and\ \bibinfo {author} {\bibfnamefont {M.}~\bibnamefont
  {Wahiddin}},\ }\href {http://books.google.ca/books?id=455\_AwAAQBAJ} {\emph
  {\bibinfo {title} {Quantum Optics for Beginners}}}\ (\bibinfo  {publisher}
  {Pan Stanford, Singapore},\ \bibinfo {year} {2014})\BibitemShut {NoStop}%
\bibitem [{\citenamefont {Nielsen}\ and\ \citenamefont
  {Chuang}(2000)}]{nielsen&chuang}%
  \BibitemOpen
  \bibfield  {author} {\bibinfo {author} {\bibfnamefont {M.}~\bibnamefont
  {Nielsen}}\ and\ \bibinfo {author} {\bibfnamefont {I.}~\bibnamefont
  {Chuang}},\ }\href {http://books.google.ca/books?id=65FqEKQOfP8C} {\emph
  {\bibinfo {title} {Quantum Computation and Quantum Information}}},\ Cambridge
  Series on Information and the Natural Sciences\ (\bibinfo  {publisher}
  {Cambridge University Press, Cambridge, England},\ \bibinfo {year}
  {2000})\BibitemShut {NoStop}%
\bibitem [{\citenamefont {Mart\'in-Mart\'inez}\ \emph
  {et~al.}(2013)\citenamefont {Mart\'in-Mart\'inez}, \citenamefont {Brown},
  \citenamefont {Donnelly},\ and\ \citenamefont {Kempf}}]{Martinez:2013}%
  \BibitemOpen
  \bibfield  {author} {\bibinfo {author} {\bibfnamefont {E.}~\bibnamefont
  {Mart\'in-Mart\'inez}}, \bibinfo {author} {\bibfnamefont {E.~G.}\
  \bibnamefont {Brown}}, \bibinfo {author} {\bibfnamefont {W.}~\bibnamefont
  {Donnelly}}, \ and\ \bibinfo {author} {\bibfnamefont {A.}~\bibnamefont
  {Kempf}},\ }\href {\doibase 10.1103/PhysRevA.88.052310} {\bibfield  {journal}
  {\bibinfo  {journal} {Phys. Rev. A}\ }\textbf {\bibinfo {volume} {88}},\
  \bibinfo {pages} {052310} (\bibinfo {year} {2013})}\BibitemShut {NoStop}%
\bibitem [{\citenamefont {Schauder}(1930)}]{Schauder}%
  \BibitemOpen
  \bibfield  {author} {\bibinfo {author} {\bibfnamefont {J.}~\bibnamefont
  {Schauder}},\ }\href {http://eudml.org/doc/217247} {\bibfield  {journal}
  {\bibinfo  {journal} {Stud. Math.}\ }\textbf {\bibinfo {volume} {2}},\
  \bibinfo {pages} {171} (\bibinfo {year} {1930})}\BibitemShut {NoStop}%
\bibitem [{\citenamefont {Ruseckas}\ and\ \citenamefont
  {Kaulakys}(2006)}]{Ruseckas:2006}%
  \BibitemOpen
  \bibfield  {author} {\bibinfo {author} {\bibfnamefont {J.}~\bibnamefont
  {Ruseckas}}\ and\ \bibinfo {author} {\bibfnamefont {B.}~\bibnamefont
  {Kaulakys}},\ }\href {\doibase 10.1103/PhysRevA.73.052101} {\bibfield
  {journal} {\bibinfo  {journal} {Phys. Rev. A}\ }\textbf {\bibinfo {volume}
  {73}},\ \bibinfo {pages} {052101} (\bibinfo {year} {2006})}\BibitemShut
  {NoStop}%
\bibitem [{\citenamefont {Kofman}\ and\ \citenamefont
  {Kurizki}(2000)}]{Kofman:2000}%
  \BibitemOpen
  \bibfield  {author} {\bibinfo {author} {\bibfnamefont {A.}~\bibnamefont
  {Kofman}}\ and\ \bibinfo {author} {\bibfnamefont {G.}~\bibnamefont
  {Kurizki}},\ }\href@noop {} {\bibfield  {journal} {\bibinfo  {journal}
  {Nature (London)}\ }\textbf {\bibinfo {volume} {405}},\ \bibinfo {pages}
  {546} (\bibinfo {year} {2000})}\BibitemShut {NoStop}%
\end{thebibliography}%

\end{document}